\documentclass[sigconf]{acmart}

\usepackage{booktabs} 
\usepackage{balance,url}
\usepackage{color,soul}
\usepackage{graphicx}
\usepackage{multirow}
\usepackage{array}
\usepackage{dsfont}
\usepackage{makecell}
\usepackage{amssymb}
\usepackage{subfigure}

\usepackage{tikz}
\usetikzlibrary{shapes, arrows, calc, positioning, fit}
\usepackage{pgfplots}
\usetikzlibrary{patterns}

\usepackage{amsmath,amssymb,amsfonts}
\usepackage{graphicx}
\usepackage{textcomp}
\usepackage{balance} 
\usepackage{algorithm}
\usepackage[noend]{algpseudocode}
\usepackage{xspace}

\newcolumntype{L}[1]{>{\raggedright\let\newline\\\arraybackslash\hspace{0pt}}m{#1}}
\newcolumntype{C}[1]{>{\centering\let\newline\\\arraybackslash\hspace{0pt}}m{#1}}
\newcolumntype{R}[1]{>{\raggedleft\let\newline\\\arraybackslash\hspace{0pt}}m{#1}}

\usepackage[colorinlistoftodos, textwidth=4cm, shadow]{todonotes}

\newcommand{\prob}{\textsc{EASS}\xspace}
\newcommand{\roprob}{\textsc{EASS-RO}\xspace}

\newcommand{\ci}{C_i\xspace}
\newcommand{\bi}{B_i\xspace}
\newcommand{\ri}{\rho_i\xspace}
\newcommand{\ef}{w_f\xspace}
\newcommand{\mf}{\lambda_f(t)\xspace}
\newcommand{\lit}{l_i(t)\xspace}
\newcommand{\lht}{\hat{l}_i(t)\xspace}
\newcommand{\sigmat}{\sigma_i(t)\xspace}
\newcommand{\alphat}{\hat{x}_i(t)\xspace}
\newcommand{\Lit}{\mathcal{L}_i(t)\xspace}
\newcommand{\xit}{x_i(t)\xspace}
\newcommand{\zit}{z_i(t)\xspace}
\newcommand{\xt}{x(t)\xspace}
\newcommand{\sit}{s_i(t)\xspace}

\copyrightyear{2020} 
\acmYear{2020} 
\setcopyright{acmlicensed}\acmConference[e-Energy'20]{The Eleventh ACM International Conference on Future Energy Systems}{June 22--26, 2020}{Virtual Event, Australia}
\acmBooktitle{The Eleventh ACM International Conference on Future Energy Systems (e-Energy'20), June 22--26, 2020, Virtual Event, Australia}
\acmPrice{15.00}
\acmDOI{10.1145/3396851.3397755}
\acmISBN{978-1-4503-8009-6/20/06}



\begin{document}

\title{Emission-aware Energy Storage Scheduling for a Greener Grid }

\author{Rishikesh Jha}
\affiliation{University of Massachusetts Amherst}
\email{rishikeshjha@cs.umass.edu}
\author{Stephen Lee}
\affiliation{University of Pittsburgh}
\email{stephen.lee@pitt.edu}
\author{Srinivasan Iyengar}
\affiliation{Microsoft Research India}
\email{t-sriyen@microsoft.com}
 \author{Mohammad H. Hajiesmaili}
 \affiliation{University of Massachusetts Amherst}
 \email{hajiesmaili@cs.umass.edu}
 \author{David Irwin}
 \affiliation{University of Massachusetts Amherst}
 \email{irwin@ecs.umass.edu}
 \author{Prashant Shenoy}
\affiliation{University of Massachusetts Amherst}
\email{shenoy@cs.umass.edu}


\renewcommand{\shortauthors}{Jha et al.}

\begin{abstract}
Reducing our reliance on carbon-intensive energy sources is vital for reducing the carbon footprint of the electric grid. Although the grid is seeing increasing deployments of clean, renewable sources of energy, a significant portion of the grid demand is still met using traditional carbon-intensive energy sources.  In this paper, we study the problem of using energy storage deployed in the grid to reduce the grid's carbon emissions. While energy storage has previously been used for grid optimizations such as peak shaving and smoothing intermittent sources, our insight is to use distributed storage to enable 
utilities to reduce their reliance on their less efficient and most carbon-intensive power plants and thereby reduce their overall emission footprint. We formulate the problem of emission-aware  scheduling of distributed energy storage as an optimization problem, and use a robust optimization approach that is well-suited for handling the uncertainty in load predictions, especially in the presence of  intermittent renewables such as solar and wind. We evaluate our approach using a state of the art neural network load forecasting technique and real load traces from a distribution grid with 1,341 homes.  Our results show a reduction of $>$0.5 million kg in annual carbon emissions --- equivalent to a drop of 23.3\% in our electric grid emissions. 

%
%
\end{abstract}

\maketitle

\section{Introduction}








A key sustainability goal of the United Nations is to attain a zero carbon economy in order to prevent 
climate change, while maintaining society's current standard of living. Doing so, involves addressing immense challenges, since it requires changing our energy consumption behavior, while also transitioning the electric grid to carbon-neutral or zero-carbon energy sources. Over the last decade, there has been an increasing deployment of clean, renewable energy sources
such as solar and wind that are already contributing positively to reducing the grid's overall carbon footprint. The levelized cost of energy from these renewable technologies is now on par or below traditional carbon-intensive generation sources, and their carbon footprint is near zero. 

However, due to their  intermittent nature, the increasing penetration of these energy sources has  increased the stochasticity and uncertainty in the grid's energy supply.
Consequently, energy storage has emerged as a related grid technology to counter this stochasticity~\cite{bnef}. Energy storage batteries can act as ``energy buffers" that smooth out the intermittent supply from renewable sources. The cost of energy storage has
continued to fall, much like that of renewables, and their deployments have begun to increase.   For instance, Green Mountain Power, a small utility in Vermont, USA, now leases Tesla Powerwall batteries to residential customers for just \$15/month, while allowing the utility to control the battery during peak periods \cite{green-mountain}.  
Such a distributed deployment of energy storage with utility control forms a type of Virtual Power Plant (VPP) that the utility can leverage for various grid optimizations~\cite{lee2018vsolar}.

Much of the recent work on energy storage-driven grid optimization has focused on \emph{demand-side optimizations} such as cost arbitrage ~\cite{walawalkar2007economics}, peak load shaving, demand response~\cite{nunna2013energy}, and ancillary services~\cite{ji2019coordinating}. Peak load shaving is a grid optimization of particular interest to utilities and involves operating  batteries during  peak demand periods in order to reduce grid stress and the reliance on peaking power plants that are operated solely to meet  peak load. Peak shaving brings economic and cost benefits, since peaking
power plants tend to be less efficient and hence the cost of supplying electricity during peak periods is much higher than at other times.  

Although energy storage-based peak shaving has been studied from a cost reduction perspective, it also brings implicit greening benefits---peaking power plants are not only less efficient  and costly to operate, they  come with a high pollution and carbon cost. Despite the implicit greening benefit from reducing the use of peaking power plants, the problem of peak load reduction using energy storage {\em does not}
directly translate to the problem of reducing the grid's carbon footprint. This is because
not all peak demand is met using ``dirty'' peaking power plants. In some cases, for instance, peak demand can be met using pumped hydro storage, which is a clean energy source, and operating energy storage batteries during such periods will not yield any emission reductions. 

Thus, reducing the grid's carbon footprint cannot be achieved by na\"{i}vely using prior methods on energy storage-based peak load reduction. This problem of emission reductions, a {\em supply-side optimization}, is not only different, but also more challenging than peak load reduction. Since  grid demand is directly observable, energy storage can be activated when peak demand occurs. Unlike observable grid demand, the grid's carbon emissions are not  directly observable  and must be inferred through other means, which is a pre-requisite for scheduling  energy storage whenever the grid's emission footprint peaks. Second, the grid is beginning to incorporate increasing amounts of clean renewable energy sources such as solar and wind, but these sources are intermittent and uncontrollable from a grid's perspective and need to be handled differently from traditional energy sources for supply-side optimizations such as emission reductions. 

The use of energy storage for explicitly optimizing the emission footprint of the grid has not been considered by prior work, with the sole exception of \cite{RPI-MS-Thesis}, where it was considered as part of a broader multi-objective optimization to reduce cost, emissions, etc., and by only considering small residential scale storage. Our work is more general since it addresses\ grid-scale storage at various levels of the grid network, and also more specific since it focuses on reducing emissions as a primary objective.
Our work is motivated by the observation that a utility typically uses a mix of generation sources to fulfill its daily demands. Different generation sources have different cost and emission footprints---for example, while coal, oil and natural gas have high emission footprints, sources such as nuclear, hydro and solar have zero emissions. The  cost of generation also varies across these generation sources. 

Utilities typically create a dispatch schedule that determines the order in which different generation sources are utilized to meet rising demand---more efficient energy sources are used more often or as base sources, while less efficient ones are often used only during high demand or peak demand  periods. 
Our insight is that these dispatch schedules and marginal analysis of energy prices can be used  to infer the carbon cost to produce the next unit of electricity at various times. Since demand is observable, we can combine this information with time-varying demand to compute the overall emission footprint at different points in the  demand curve and then intelligently activate energy storage whenever the emissions footprint is high; as noted, the emissions footprint depends on the energy fuel sources used   and not on the demand, yielding a different schedule for operating energy storage than that from peak shaving. Such emission-aware scheduling of energy storage can provide significant benefits in greening the grid.

However, there are many challenges in designing algorithms for emission-aware scheduling of 
energy storage. First, the daily electricity demand is stochastic and time-varying and also depends on weather conditions. Second, as the penetration of ``clean" renewables such as solar and wind increases, it naturally lowers the emission footprint of the grid but also increases uncertainty and stochasticity in demand due to the net-metering of these intermittent sources. Third, the emission footprint of various sources, represented by the marginal cost of generation, is itself time-varying due to changes in prices of inputs and other factors.
Finally the energy storage deployment will be distributed and heterogeneous with batteries of various sizes and technologies deployed in different parts of the grid. 

In this paper, we leverage robust optimization~\cite{ben2009robust,bertsimas2004price} (RO) to tackle the uncertainty of the daily electricity demand. Classic stochastic optimization approaches require stochastic modeling of uncertain parameters, and deviations from the models may degrade the performance of the proposed solutions substantially. In contrast, RO does not rely on an underlying probability distribution of the uncertain input, and only requires limited information of the uncertain data, including mean, and interval predictions, i.e., upper and lower bounds of the uncertain data. When compared to probability distributions, mean and interval prediction values are much simpler to  estimate. In addition, RO always calculates a solution that is \textit{guaranteed} to be feasible within all possible realizations of the uncertainty sets. Note that RO and competitive algorithm design~\cite{Borodin98} are two approaches in the literature that do not rely on any stochastic modeling of the uncertain data. While a competitive approach is too conservative since it guarantees worst-case performance, the additional interval prediction data can result in better performance in RO. 
In designing and evaluating our emission-aware storage scheduling approach, our 
paper makes the following contributions.\\
\textbf{Problem formulation.} We present a detailed formulation of the problem of emission-aware
storage scheduling as an optimization problem. Our formulation is sufficiently general
to incorporate a range of possibilities, including heterogeneous storage deployments,  
distributed renewable generation, and  time-varying marginal costs. 
\\
\textbf{Emission-aware  energy storage scheduling using robust optimization.}
 We use robust optimization, a stochastic optimization approach, to solve the energy-aware
 storage scheduling problem. As noted above, the use of robust optimization allows us to find a solution that is guaranteed to be feasible within all the possible realizations of the input in a predetermined uncertainty set. 
\\
\textbf{Load forecasting under uncertainty.} Since our optimization
approach requires load predictions, we also use a state of the art autoregressive neural network algorithm for transformer load forecasting, which is then utilized by our optimization approach. Our forecasting method can handle uncertainty in demand from stochastic time of day effects as well as that from net-metered renewables. 
\\
\textbf{Grid-scale evaluation.} We present a grid-scale evaluation of our approach
using real traces from a distribution grid comprising 100 transformers and 1,341 homes. Our results show carbon emissions savings of $>$0.5 million kg over a period of a year. This reduction 
is equivalent to 23.3\% of overall emissions from the electric grid. We also show that even at 50\% storage penetration level we can achieve up to 13.9\% reduction in carbon emissions.

\section{Background\label{sec:back}}

In this section, we provide background on electric grids, generation sources, and energy storage.


\subsection{Electric grids}

\begin{figure*}[t]
	\centering
	\includegraphics[width=0.9\textwidth]{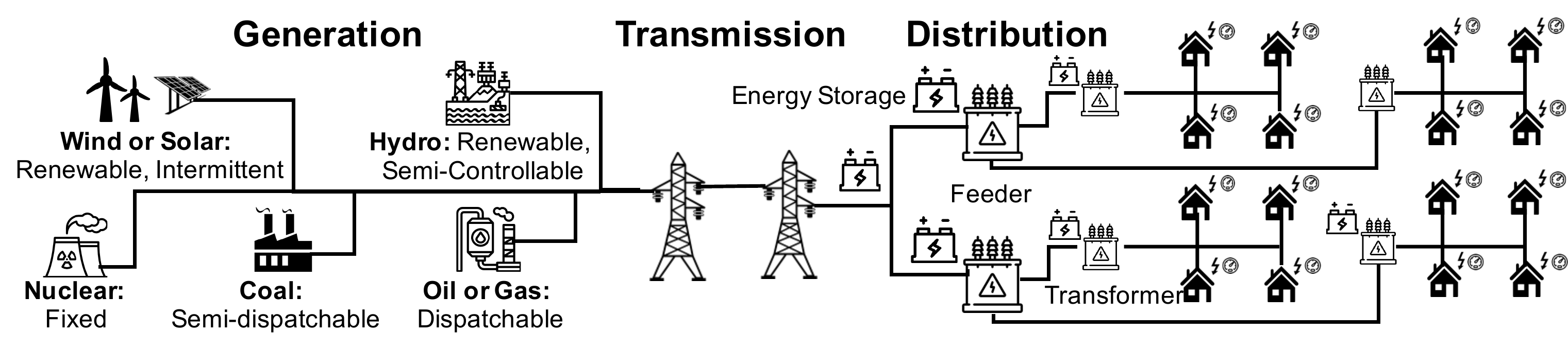}  
	\caption{Electric grid architecture comprising generation, transmission and distribution. Our work assumes a heterogenous energy storage within the distribution grid. }
	\label{fig:scenario}
\end{figure*}

As is well known, today's electric grids comprise three components: generation, transmission, and distribution (see Figure~\ref{fig:scenario}). A key goal of the grid operator is to ensure that demand and supply are matched at all times for proper functioning of the grid.  Since electricity demand changes continuously over the course of a day (see Figure~\ref{fig:data_sample}(a)), the generation must be matched to changing demand via a  \textit{dispatch schedule}~\cite{zhang2018peak}. The dispatch schedule determines the order and schedule for activating and deactivating various generators that are at the disposal of the grid operator and specifies how the supply should be ramped up or down to match time-varying demand. 

Typically, power plants and generators used for the dispatch schedule fall into
three categories: (i) \textit{Base load generators}: These are generators at power plants
that operate at all times to support the base demand; generators at large-scale power
plants such as nuclear, natural gas, coal, and biomass fall into these categories;
(ii) \textit{Load following generators:} These are generators that are activated during the high demand periods within each day (such as morning and evening) to support demand
beyond the base load; (iii) {\em Peaking generators:} These are standby generators that 
are activated when the overall demand hits seasonal peaks. They may operate for only a few
days of the year when the hot or cold weather causes the demand to peak for the season.
In general, peaking generators tend to be older, less-efficient generators within the overall mix that are kept on standby for infrequent use; old coal and oil generators that are nearing the end of their lifetime are examples of peaking generators.  Note that
high or peak demands can also be met through other means, such as pumped hydro storage, and hence, {\em the emission footprint may not always  rise with  demand.} 


In contrast to traditional sources of electricity generation, renewable sources such as solar, wind and hydro are non-polluting in nature and have zero carbon emissions. 
Distributed renewable energy sources such as solar tend to be part of the distribution network and often net-meter their power output directly into the distribution grid. Further, renewable sources such solar and wind are assumed to be uncontrollable due to their intermittent nature and thus not dispatchable.  

\subsection{Emission from Generation Sources}
As discussed below, the carbon intensity, and the resulting emission footprint, of the grid varies continuously over the course  of the day. If the emission footprint from generation were  directly observable, we could simply schedule energy storage whenever emissions peak during each day. Since the emission footprint is not directly observable, we need to infer it through other means for our emission-aware scheduling approach. 
Two factors need to be considered for doing so: the \textit{average carbon intensity} and \textit{marginal carbon intensity}. 


The \textit{average carbon intensity} of an electric grid is defined as the weighted average of \textit{emission factors} of the available fuel types, in which 
\textit{emission factor} of each fuel type is defined as its carbon emission by generating one unit of electricity. Table~\ref{tbl:carbonintensity} lists the values of emission factor for the available generation types in ISO New England~\cite{emissionfactor}. The average carbon intensity is the weighted average of emission factors for the energy mix used by the grid. For example, if an electric grid produces electricity from coal, natural gas, nuclear, and hydro in equal proportions, then the average carbon intensity would be 339.49 kg/MWh ($962.97\times0.25 + 395.53\times0.25 + 0\times0.25 + 0\times0.25$), in the above example. 

In general,  however, the reduction or increase in generation, and consequently, carbon emissions, due to changes in electric demand  (dictated by the dispatch schedule) is not the same across all power plants.
Most of the changes occur in the load following power plants, and occasionally in the peaker power plans, which we collectively refer to as \textit{marginal power plants}. 
These are generators that can be ramped up or down at short notice to respond to changes
in demand as determined by the dispatch schedule. Consequently, when attempting to reduce
the emissions from the generation mix, we must consider the \textit{marginal carbon intensity}, which is the emissions from generating the next unit of electricity; in our  case, it is related to the operation of marginal power plants.
In the above example, if the marginal power plant uses natural gas as fuel, the marginal carbon intensity is 395.5 kg/MWh, which is higher than the average carbon intensity in our example.  Since there are notable differences in the emissions factors associated with the different fuel types, there is significant potential for reducing the carbon footprint of the  overall electricity generation by optimizing the marginal carbon intensity through the use of energy storage.

\begin{figure*}[t]
\centering
\begin{tabular}{cc}
\includegraphics[width=2.5in]{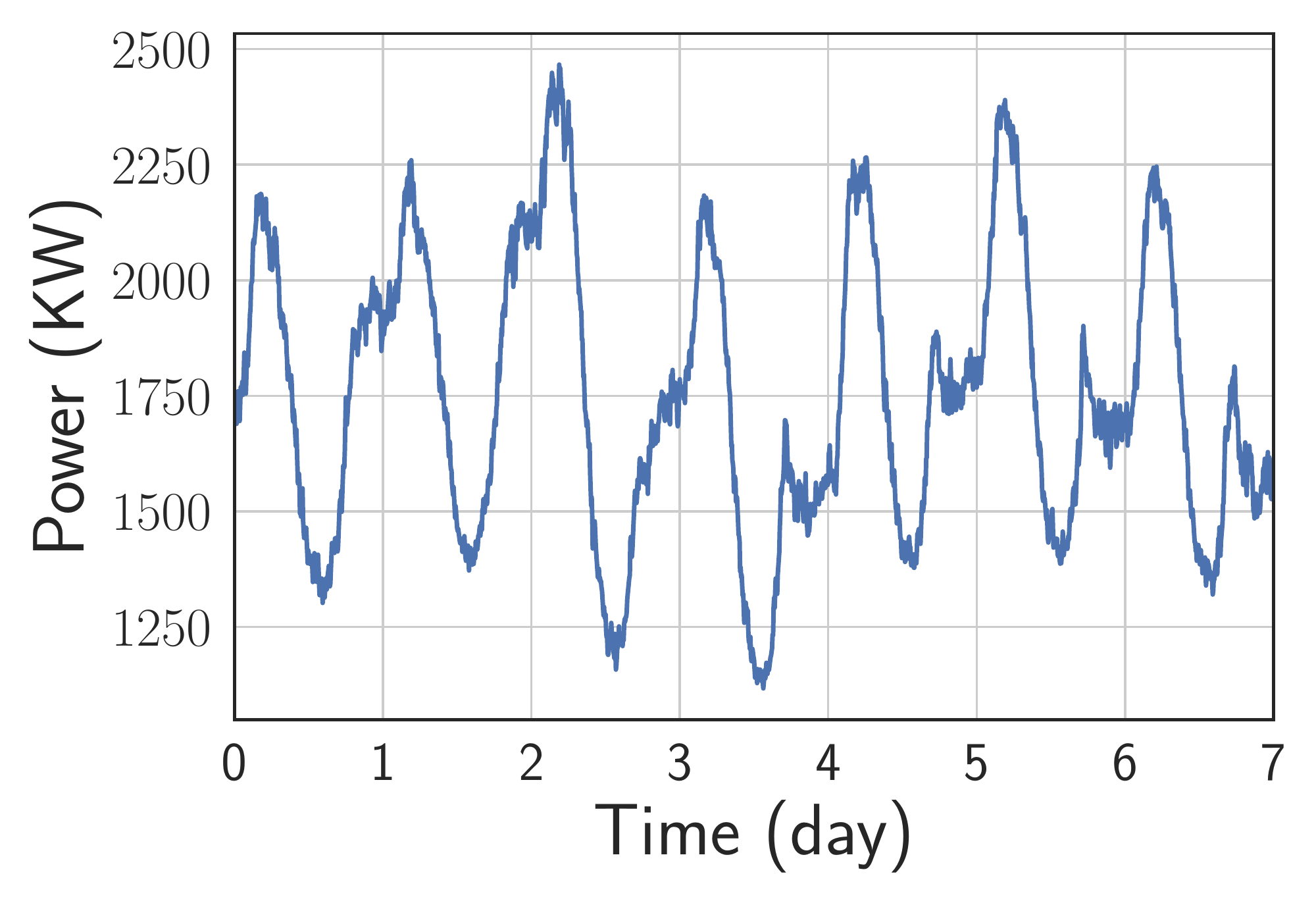} &
\includegraphics[width=2.5in]{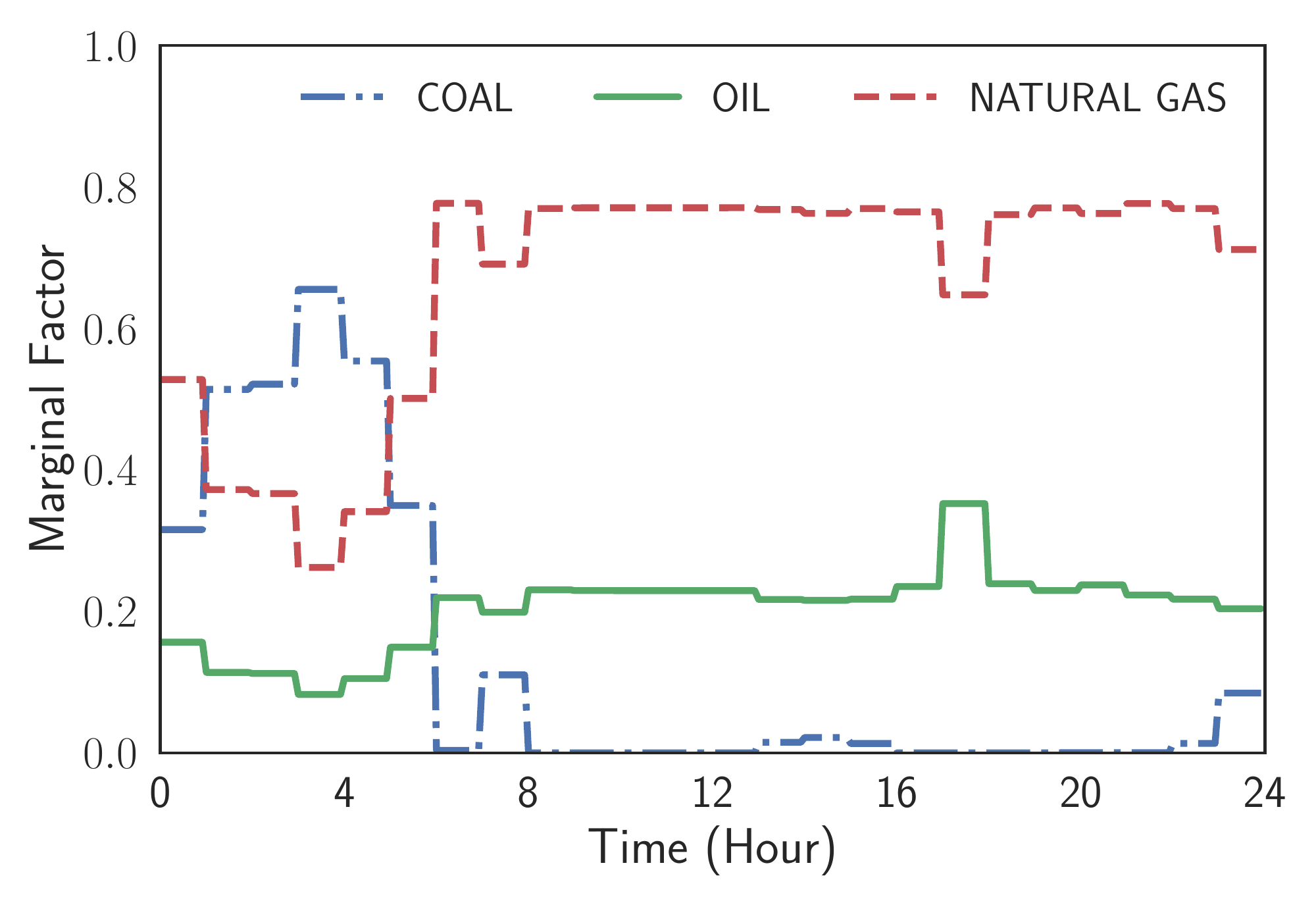}  \\
(a)  & (b) \\ 
\end{tabular}
\caption{(a) Daily transformer load (b) Marginal factor of different fuels for a day.}
\label{fig:data_sample}
\end{figure*}

It should also be noted that the marginal carbon intensity will vary over time due to several factors. For example, if the generation from hydro plants has to be decreased during periods of little rain, generation from other (less green) sources will have to make up the shortfall, potentially increasing the marginal carbon intensity.  Fuel prices of raw materials such as natural gas and oil may fluctuate over time, and dispatch
schedules may be optimized to use cheaper sources. The dispatch schedule itself varies over the course of a season based on seasonal demand. All of these factors cause the marginal carbon intensity to vary, and any approach that seeks to optimize marginal
emissions must account for such temporal variations.  Figure~\ref{fig:data_sample}(b) illustrates the variations of the marginal carbon intensities of different fuel types.

\begin{table}[]
\centering
\caption{Carbon emission by different generation types, data publicly available from~\cite{emissionfactor}}
\label{tbl:carbonintensity}
\begin{tabular}{|c|c|}
\hline
\textbf{Generation Type} & \textbf{Emission Factor (CO$_{2}$ kg/MWh)} \\ \hline \hline
Coal                                                       & 962.97                                                                  \\ \hline
Natural Gas                                                & 395.53                                                                   \\ \hline
Oil                                                        & 933.94                                                                  \\ \hline
Nuclear                                                    & 0                                                                     \\ \hline
Hydro                                                      & 0                                                                     \\ \hline
Solar and Wind                                             & 0                                                                     \\ \hline
\end{tabular}
\end{table}


\subsection{Renewable Energy Sources} 
The penetration of renewable sources such as solar and wind in the grid has grown substantially in recent years. These clean sources have zero carbon emissions and directly contribute to a reduction in the grid's overall carbon footprint. However, the generation from renewable sources is known to be intermittent and dependent on the weather. As a result, the grid treats these generation sources differently from traditional sources such as natural gas and coal. In particular, today's grid operators assume that these sources are uncontrollable, and hence {\em not dispatchable}. That is, their generation potential at any instant in the future is not entirely predictable, and they are typically not included into dispatch schedules unlike other sources of energy generation. Instead, their output is assumed to be net-metered into the distribution grid, which means that renewable sources are treated as part of the fluctuating demand, rather than as an explicit dispatchable supply source. In line with the grid's assumptions, our paper also assumes that renewables such as solar and wind are handled as a portion of stochastic demand, and increasing penetration results in higher stochasticity and uncertainty in demand. 

\subsection{Energy Storage}

Grid energy storage technologies in the form of batteries have been gaining traction in recent years. Companies such as Tesla have deployed both small- and large-scale energy storage within the grid in many different countries and demonstrated the feasibility and benefits of using such storage for grid optimizations. This work assumes a distributed deployment
of energy storage batteries within the distribution grid. The deployment is assumed to
be heterogeneous --- the sizes of batteries and the level within the grid where they are deployed are assumed to be different for different batteries. Some batteries may be small batteries, akin to the Tesla Powerwall, deployed adjacent to small neighborhood distribution transformers. Other batteries may be larger in size and deployed near larger transformers (potentially at the feeder or substation levels) that supply electricity to a larger number of homes.  The penetration of energy storage within the distribution grid can differ from one scenario to another, and our work is designed
to handle different penetration levels.

The distributed network of batteries is assumed to be under the control of the grid operator. 
However, rather than using them for grid optimizations, our work seeks to operate this distributed
set of batteries to minimize the aggregate carbon emissions of the grid, given the time-varying marginal carbon intensity values---by operating them to reduce reliance on the dispatchable sources with high marginal carbon intensities. We formulate this problem of emission-aware scheduling of energy storage more formally in the next section.

\section{Problem Formulation\label{sec:prob}}

Consider a distribution grid comprising a network of substations, feeders and neighborhood
transformers. Assume that energy storage is deployed at various points within this grid, i.e., at a subset of the transformers. 
In a typical grid, there is significant variation in the capacity of transformers and the number of customers it serves. Thus, the energy storage must be sized according to the transformer capacity to enable grid optimizations at that location. Further, our work assumes that this heterogeneous collection of batteries  is under the control of the grid operator and the operator can control the charging and discharging of a distributed network of batteries.  

As discussed in the previous section, there are several fuel types in a typical electric grid, each with a different level of emission factor and different  time-varying marginal carbon intensity. The emission footprint of the grid at any instant depends on the mix
of generation fuels used to satisfy the current demand. Grid operators must match supply and demand at all times by constructing  a  {\em dispatch schedule} in advance. The dispatch schedule is typically computed a day in advance by first estimating the demand curve for the following day and then determining an order in which different generation sources are  activated (or deactivated) to meet the predicted rise and fall in demand.\footnote{As noted earlier, renewable sources such as solar and wind are assumed to net-metered into the distribution grid and not considered as dispatchable energy resources.}
This problem  becomes a cost minimization problem since the next unit of rise in demand should be satisfied using the generation source with the cheapest marginal price \cite{zhang2018peak}. This problem  is also referred to as the unit commitment problem  in power system literature ~\cite{Padhy04}; in solving  this dispatch schedule (i.e., unit commitment), time-varying marginal prices are computed at each step
to select the least cost  source for each unit change in demand.  Since this problem is solved one day in advance (usually through day-ahead energy markets), we 
assume that time-varying marginal factors can be obtained when the dispatch schedule is finalized at  the start of each day.  

The key insight  behind our approach is to take the 
demand seen within the distribution grid (which is directly observable)  and these computed marginal prices at  different times of the day to infer the emission footprint of the grid over time; the scheduling problem is then to intelligently schedule the batteries during peak emissions periods, subject to various demand constraints within the distribution grid. This leads us to the following problem:
\textit{Given the dynamics in the marginal factor of available fuel types and in the distribution-level demand, what is the optimal scheduling of energy storage that minimizes grid-wide carbon emissions and respects the operational constraints of the grid and energy storage systems?}

\begin{table}[t]
	\caption{Summary of notations}
	\label{tbl:not}
	\begin{center}
		\begin{tabular}{|c|L{6.8cm}|}
			\hline
			\multicolumn{2}{|c|}{\textbf{Inputs}}\\
			\hline\hline
			$T$ & The number of time slots, $T\geq1$\\
			$\mathcal{T}$ & Set $\mathcal{T} = \{1, 2, \dots, T\}$\\
			$n$ & The number of transformers\\
			$m$ & The number of fuel types\\
			$\ci$ & The capacity of transformer $i$\\
			$\bi$ & The capacity of storage system at transformer $i$\\
			$\ri$ & Charge and discharge rate limit of storage system $i$\\
			$\ef$ & Emission factor of fuel type $f$ \\
			$\mf$ & marginal factor of fuel type $f$ at $t$\\
			$\lit$ & Mean value of day-ahead forecast load at transformer $i$ at $t$\\
						\hline\hline
			\multicolumn{2}{|c|}{\textbf{Optimization variables}}\\
			\hline
			\hline
			$\xit$ & The charge/discharge amount of storage $i$ at $t$\\
			$\xt$ & The aggregate charge/discharge at $t$, i.e., $\xt = \sum_{i = 1}^{n} \xit$\\
			$b_i(t)$  & The storage level (state of charge) of storage $i$ at the end of $t$\\

			\hline
		\end{tabular}
	\end{center}
\end{table}

\subsection{System Model}
In this section, we formulate the \textit{offline} version of the emission-aware storage scheduling problem (\prob) assuming that the entire load data is available in advance, and in the next section, we present the \textit{online} formulation that takes into account the uncertainty of load.

We assume that the time horizon is divided into $T$ real-time settlement intervals, indexed by $t$, each with fixed length. Time slots are set according to the real-time settlement intervals in the U.S.-based electricity markets, e.g., $5$ minutes in CAISO and NYISO, and $15$ minutes in ERCOT~\cite{epri2016}. The main notations are summarized in Table~\ref{tbl:not}. In what follows, we explain the details of the system model.

Assume there are $n$ transformers in the system, each indexed by $i$. 
Further assume that there is an energy storage battery at each transformer.  In practice,
the operator may only deploy energy storage at a subset of transformers, which can be easily modeled by setting the sizes of batteries at all other transformers to zero.
The scheduling decisions are assumed to be made at the transformer level. Let $\ci$ be the capacity of transformer $i$, and $\bi$ be the storage capacity deployed at transformer $i$. Let $\ri$ be the maximum charging and discharging rate of storage $i$. Our formulation could be extended to the case with different charge and discharge rate constraints. 

We assume there are $m$ different fuel types in the grid, each indexed by $f$. Let $\ef$ be the emission factor in kg/MWh of fuel type $f$. Further, let $\mf$ be the marginal factor, as the contribution of  fuel type $f$ at time $t$ to the marginal increase or decrease in energy demand. The values of $\ef$ are fixed and given in advance. The values of $\mf$ change over time based on solving grid dispatch and unit commitment problems. 

Finally, let $\lit$ be the day-ahead load forecast of transformer $i$ at time $t$. 
In addition, let $\lht$ be the actual values of load in real-time. Note that $\lit$ and $\lht$ might be different since there is always some error between the forecast day-ahead and actual values. The problem formulation in this section is an \textit{offline} version that takes into account the day-ahead load values. In Section~\ref{sec:ro}, we extend the formulation to include the uncertainty of actual load in real time.

{\noindent \bf Incorporating renewables:} Since renewables such as solar and wind are assumed to net-metered into the distribution grid, we assume that the transformer-level demand $\lit$ is a net-metered value that represents the difference between the actual demand at that transformer and the generation from any renewable sources, such as rooftop solar, that are present at that location. This has two implications. First, it allows our approach to naturally incorporate the contribution of renewables for emission reduction (since it contributes to a direct reduction in demand). Second, it increases the stochasticity in the demand, making demand more unpredictable, an issue we must address as part of our optimization.

\subsection{Problem Formulation}
\paragraph{The Optimization Variables} The charging/discharging amount of storage $i$ at $t$ is represented as $\xit$. Positive values, i.e., $\xit > 0$ indicate the charging of the storage, whereas negative values, i.e., $\xit < 0$ indicate discharging. The aggregate change in load due to charge/discharge of different storages observed at the grid level at time $t$ is represented by $\xt$, i.e., $\xt = \sum_{i = 1}^{n} \xit$.
Finally, let $\sit$ be the state of charge of storage $i$ at time $t$, and we will obtain its evolution over time in the following by formalizing the constraints.
\\
\paragraph{Constraints} The change in load observed at the grid level is the sum of all the charging/discharging decisions made at each transformer, i.e., 
\begin{equation}
\label{eq:xt}
\xt = \sum_{i = 1}^{n} \xit, \quad \forall t,
\end{equation}
The evolution of storage is represented as the following constraint
\begin{equation}
s_i(t+1) = \sit + \xit,   \quad  \forall t \text{ and } \forall i,
\end{equation}
The other constraints regarding the physical limits of storage are presented as follows. The scheduling decisions should be taken within the operating constraints of the battery. For example, the maximum charge of the storage unit should not exceed the storage capacity while it is in operation. For simplicity, we assume the maximum charging and discharging rate to be equal. We represent the constraints regarding the state of the charge of storage as follows
\begin{eqnarray}
\label{eq:cap}
&& \sit \leq \bi, \quad  \forall t  \text{ and } \forall i,\\
\label{eq:str_neg}
&&\sit \geq 0, \quad \forall t  \text{ and } \forall i,\\
\label{eq:str_rate}
&&-\ri \leq \xit \leq \ri, \quad  \forall t  \text{ and } \forall i.
\end{eqnarray}

In order to maintain the demand and supply relationship, the discharge from the storage unit should not be greater than the load observed at the transformer at any time $t$. Thus, we have
\begin{equation}
\label{eq:xtllt}
- \xit \leq \lit, \quad  \forall t \text{ and } \forall i.
\end{equation}
Note that in reality, it is possible to have peak load beyond the capacity of transformers. In case of the load at the transformer at time $t$ is greater than the capacity of the transformer, we do not want to worsen the situation by charging the storage at that time leading to transformer overload. Hence, by defining parameter $\eta$ as the threshold violation level of transformer capacity, we express the this constraint as 
\begin{equation}
\label{eq:neg_xt}
\ci - \xt - \lit \geq \eta, \quad \forall i \text{ and } \forall t.
\end{equation}
The value of $\eta$ is set by the operator of the grid, and in experiments we set it to 1\% of the transformer capacity. 
\paragraph{Objective Function} The eventual goal is to minimize the carbon emission of the grid by managing the charging and discharging of the energy storage. 
More specifically, we aim to minimize the following objective function: 
\begin{equation}
\sum_{t=1}^{T} \sum_{f=1}^m \ef \mf \xt,
\end{equation}
where $\xt$ represents the aggregate change in load by scheduling the storage observed at grid level as described in~\eqref{eq:xt}. Recall that $\mf$ represents the marginal factor of fuel type $f$ at time $t$, and $\ef$ is the emission factor of fuel type $f$.

\paragraph{Optimization Problem Formulation}
Putting together, we formulate the emission-aware energy storage scheduling (\prob) problem as 
\begin{eqnarray*}
\prob: &\min & \sum_{t=1}^{T} \sum_{f=1}^m \ef \mf \xt\\
&\text{subject to:} & \text{Equations}~\eqref{eq:xt}-\eqref{eq:neg_xt},\\
&\text{variable:} & \xit \in \mathbb{R}, i \in \{1,\dots,n\}, t \in \{1,\dots,T\}.
\end{eqnarray*}

The \prob problem is linear in nature that could be solved optimally if the entire input to the problem, i.e., load values and emission parameters, are given in advance. In practice, however, these values are uncertain, and  as we will show in Section \ref{sec:forecast}, future predictions of load are never 100\% accurate. Consequently, in the next section, we introduce the robust optimization formulation to tackle the uncertainty that arises when solving this problem \emph{online} in real-world settings.

\section{Robust Optimization Approach\label{sec:ro}}

In this section, we present the robust optimization version of \prob (called \roprob) by taking into account the uncertainty due to the imbalance between the forecast and actual real-time load values. 
Robust Optimization (RO)~\cite{bertsimas2004price} is a well-established framework for general scenarios of decision making under uncertainty. In this paper, we leverage the RO framework for emission-aware storage scheduling under the uncertainty of electricity load. 
As compared to the traditional stochastic optimization approaches, problems formulated in an RO framework are typically computationally tractable and do not require the knowledge of a probability distribution over the uncertain input. 

The first challenge in formulating an RO counterpart of \prob is to define an \textit{uncertainty set} which bounds the upper and lower bounds that the uncertain input, i.e., load,  can take. The classic approach in RO is to optimize for the worst case value in the uncertainty set, but that might be too conservative leading to a suboptimal solution. 
In this paper, we follow another variant of RO framework, called the price of robustness, proposed in~\cite{bertsimas2004price}. More specifically, Bertsimas, et. al.~\cite{bertsimas2004price} develop a generic uncertainty set that can be used to formulate a robust linear counterpart of an uncertain linear program. In this approach, the level of \textit{robustness} can be controlled by parameter $\Gamma$ known as the \textit{budget of uncertainty}. Then they proved that $\Gamma$ can be chosen based on the level of robustness desired by the operators such that the probability that the constraint is satisfied is $1-\epsilon$.  In what follows, we present the robust counterpart of the \prob problem using the robust framework in~\cite{bertsimas2004price}, 

\paragraph{Robust Counterparts of Uncertain Linear Constraints}
In this section, we present the robust counterparts of the constraints that include load values. 

First, we state the robust counterpart of constraint~\eqref{eq:xtllt}. 
Recall that constraint~\eqref{eq:xtllt} is enforced to ensure that discharge from the battery should be less than the load observed at the transformer. 
The detailed steps toward stating constraint~\eqref{eq:xtllt} in a robust framework is the following. 
First, we construct the uncertainty set associated with the transformer level loads. 
The uncertainty set for actual load of transformer $i$ at time $t$ can be represented as 
\begin{equation}
 \Lit = [\lit - \sigmat,\lit + \sigmat],
\end{equation}
where $\sigmat$ is the deviation from the expected value. 
The values $\sigmat$ and accordingly $\Lit$ should be obtained by using a forecast model of the transformer level load. In Section~\ref{sec:load_forecasting}, we present our forecasting approach based on state-of-the-art neural networks for predicting the load. The values of $\sigmat$ will be used to construct the robust constraint.

By defining $\Gamma$ as the \textit{budget of uncertainty}~\cite{bertsimas2004price}, we re-express constraint~\eqref{eq:xtllt} as 
\begin{equation}
    -\xit - \lit +  \beta_i(\lit, \Gamma) \leq 0 \quad  \forall i \text{ and } \forall t,
\end{equation}
where $\beta_i(\lit,\Gamma)$ represents deviation of $\lit$ from its expected value, given $\Gamma$ as the budget of uncertainty. Note that $\Gamma$ could be readily extended to be defined for each transformer separately. In the following, we explain how to calculate the value of $\beta_i(\lit,\Gamma)$.

In the original robust optimization framework under the paradigm of price of robustness~\cite{bertsimas2004price}, the budget of uncertainty is defined for each uncertain constraint separately, and its goal is to provide a trade-off between the robustness against the performance of the solution. More specifically, the value of $\Gamma$ determines that for each constraint how many elements should be robust against violation; the higher the value of $\Gamma$, the higher the robustness, the lower the performance. In other words, the solutions with the higher values of $\Gamma$ might be suboptimal since it is too conservative for the sake of  ensuring robustness. 

Since constraint~\eqref{eq:neg_xt} is independent for each transformer load at each time slot, following the original approach in~\cite{bertsimas2004price} requires us to have $n\times T$ separate values robust constraints each for one instance of~\eqref{eq:xtllt}. This approach does not provide any flexibility to determine the level of robustness and limits us to the case with the maximum robustness in solution. To provide flexibility for trade-off between robustness and performance, we slightly change the original framework by considering a common budget of uncertainty for the entire time horizon of each transformer, i.e., grouping all the constraints of each transformer over time.

More specifically, let $\boldsymbol{z}^{\star} = [z_i^{\star}(t)]_{t\in \{1,2,\dots, T\}}$ be the optimal value of the following optimization problem~\cite[Section 3, Proposition 1]{bertsimas2004price}:
\begin{eqnarray}
\boldsymbol{z}^\star = \arg \max_{\zit \in   [0,1]} \sum_{t=1}^T \sigmat  \zit, \quad \text{s.t.}\quad  \sum_{t=1}^T \zit \leq \Gamma.
\end{eqnarray}
Note that the above problem should be solved for each transformer $i$ separately. Then, we calculate the values of each $\beta_i(\lit, \Gamma)$ as follows:
$$\beta_i(\lit, \Gamma) = \sigmat z_i^{\star}(t).$$

Using this formulation one can see that $\beta_i(\lit, \Gamma)$ ranges between $[0,\sigmat]$, thus $\lit$ ranges between $[0, \lit + \sigmat]$. We can safely ignore the set $[\lit - \sigmat, \lit]$ as we want to make our solution robust to the worst case scenario. Using the budget of uncertainty parameter $\Gamma$ we can control the level of robustness across time in terms of the following optimization problem. 

Intuitively, we can see how the value of $\Gamma$ controls the deviation from $\lit$. In case we set $\Gamma = 0$ the \roprob problem reduces to \prob as all the deviations from $\lit$ will be 0. On the other hand, increasing $\Gamma$ increases the deviation of $\lit$ thereby we have more  robustness in the solution. When $\Gamma = |T|$ each $\zit$ will be set to 1, thus \roprob is the most conservative formulation. Last, in \prob, constraint~\eqref{eq:neg_xt} is involved with the uncertainty of the load. Hence, the same procedure as for constraint~\eqref{eq:xtllt} should be done to have its robust counterpart.

%
%


\subsection{Load Forecasting Under Uncertainty}
\label{sec:load_forecasting}

Our approach requires load forecast as input and internally deals with its uncertainty. While there are have been several research on forecasting demand~\cite{kyriakides2007short}, most approaches focus on predicting the aggregate grid demand, which is often smooth and predictable. However, transformer load sees higher variations depending on the number of homes the transformers feed electricity~\cite{iyengar2016analyzing}.  Further, any net-metered renewable sources such as rooftop solar or wind will increase the stochasticity in the observed demand.  As such, it is more challenging to provide accurate forecasts and has higher uncertainty in prediction, which justifies the need for the use of robust optimization methods that can handle such uncertainty. Formally, forecasting transformer loads requires learning a function $F_i$ that predicts future loads based on input parameters stated as follows  
\begin{align*}
l_i(t+1,t+2,\dots,t+k) 
= F_i(l_i(1,2,\dots,t), \tau) \quad \forall i.
\end{align*}
where $F_i$ predicts future load for the next $k$ time steps, and $\tau$ is a vector that represents exogenous feature inputs such as temperature and  day of the week. 

Load at the transformer level shows both diurnal and weekly patterns~\cite{iyengar2016analyzing}. For example, load during mid-day will differ from load seen at night. Similarly, weekday load differs from weekend load patterns. 
Although transformer level load data is noisy in nature,  historical load data contains daily and weekly seasonality. It is important to extract the seasonality in the historical data for accurate forecasting. We use this insights to model our load. 
Specifically, our approach is based on an \emph{Autoregressive Neural Network}~\cite{diaconescu2008use}. 

In order to forecast load for time $t + 1$ we use the past $p_1$ time slot as the input along with loads on past $p_2$ days at time $t + 1$ as well as loads at past $p_3$ weeks at time $t + 1$.
Along with the historical load, we include one-shot encoded day of the week exogenous variable as part of feature vector of the neural network. We also use the temperature $\tau_{t+1}$ at time $t + 1$ as an external regressor. Putting together all above inputs, the forecast load at time $t+1$ is as follows
\begin{align*}
l_i(&t+1) =\beta \tau_{t+1} \\
 + &\texttt{N}\Big(\delta, l_i(t), l_i(t-1),\dots, l_i(t-p_1),\\
   & l_i(t + 1 - T_{\textsf{d}}, t + 1 - T_{\textsf{d}}\times 2, \dots, t + 1 - T_{\textsf{d}}\times p_2), \\
   & l_i(t + 1 - T_{\textsf{d}}\times 7, t + 1 - T_{\textsf{d}}\times 7\times 2, \dots, t + 1 - T_{\textsf{d}}\times 7\times p_3)\Big),
 \end{align*}
 where $T_{\textsf{d}}$ is the number of time slots in one day. In our experiments, the length of each slot is 5 minutes, hence, $T_{\textsf{d}} = 12 \times 24 = 288$.

\section{Evaluation Setup\label{sec:setup}}
In this section, we discuss our experimental setup and methodology.

\subsection{Experimental Datasets}
\label{sec:dataset}
\paragraph{Load Dataset}
For evaluating the efficacy of our load forecasting techniques along with the distributed storage schedule, we use a grid-scale dataset obtained from an utility company in the Northeastern US containing energy data from 1,341 smart meters connected to 100 transformer. This data is available at a 5-minute granularity over a period of 2 years. On an average, each transformer is connected to 13.4 smart meters (ranging from 5 to 85).
Likewise, the transformer capacity varies  between 25 and 750 kVA.  Table \ref{tbl:dataset} summarizes our dataset.

\begin{table}[t]
\caption{Dataset used for our evaluation}
\label{tbl:dataset}
\begin{tabular}{|p{4.7cm}|p{3.2cm}|}
		\hline
		\textbf{Characteristics}                              & \textbf{Value}                  \\ \hline
		Number of homes & 1,341           \\ \hline
		Transformers                 & 100 \\ \hline
		Transformer size                & 25 to 750 kVA \\ \hline
		Trace Duration              & 2 years      \\ \hline
\end{tabular}
\end{table}

\paragraph{Marginal Carbon Intensity}
Additionally, our 
scheduling scheme requires marginal carbon intensity as the input to the problem. This data is not directly available for the New England region. Hence, we use the method specified in~\cite{rogers2013evaluation} that estimates the marginal power plants in operation using the hourly locational marginal price (LMP) of electricity generation and the monthly fuel prices available through~\cite{eia} and \cite{iso-ne}. 
The approach uses a symmetric Gaussian membership function in (\ref{eqn:member}) that maps the LMP values to fuel types.
\begin{equation}
	M_f(t) = e^{\frac{-(p(t) - \mu_f)^2}{2\nu_f^2}}
	\label{eqn:member}
\end{equation}
where $\mu_f$ and $\nu_f^2$ is the average cost and variance of fuel type $f$, and $p(t)$ is LMP of the market at time $t$.
Subsequently, we compute the marginal factor $\lambda_f(t)$ of fuel type $f$ as follows.
\begin{equation}
\lambda_f(t) = \frac{M_f(t)}{\sum_f M_f(t)}
\end{equation}

\subsection{Experimental Settings and Baseline}
\paragraph{Parameter Settings} 
We set the time horizon to one day, and the length of each slot is 5 minutes, hence $T = 12\times24 = 288$. We initialize the storage level at half its total capacity to allow both charging and discharging starting with time $t = 0$. We also constrain the storage capacity at the end of the day to half its capacity so as to have the same state of charge for the next day, i.e., 
\begin{equation}
\sit = \frac{\bi }{2}; \quad \text{ if } t = 1 \text{ and } t = T = 288,  \forall i. \nonumber
\end{equation}

While evaluating our robust approach, we use load values and fuel type parameters directly read from the dataset described above. The additional parameters are described in Table~\ref{tbl:linearprogramming}. 

\begin{table}[]
\caption{Parameter settings of our approach.}
\label{tbl:linearprogramming}
\begin{tabular}{|p{4.7cm}|p{3.2cm}|}
		\hline
		\textbf{Parameters}                              & \textbf{Value}                  \\ \hline
		Charge/Discharge Rate Limit & 60 mins           \\ \hline
		Marginal Fuel Sources                   & Coal, Oil, Gas \\ \hline
		Emission Factor (kg/MWh)               & Refer to Table 1       \\ \hline
		$\eta$ in Equation~\eqref{eq:neg_xt}&  1\% of $C_i$ \\ \hline
		$\Gamma$ in \roprob & [10,20] \\\hline
	\end{tabular}
\end{table}

\subsubsection{Baseline Algorithms}
\label{sec:baseline}
We compare the performance of \roprob with the following approaches.
\begin{enumerate}
	\item \textit{Optimal Offline Solution:} The optimal offline approach assumes complete knowledge of future load and provides the best achievable schedule to minimize carbon emissions. Although not practical, it serves as a best theoretical upper bound to compare with.  
	\item \textit{Online Linear Programming:} In this approach, we use the day-ahead forecast load as input to solve the linear program \prob and determine the schedule. However, the day-ahead charge/discharge schedule may violate real-time grid constraints as the actual load at time $t$ may differ from predicted load at time $t$. To ensure that all grid constraints are satisfied, the day-ahead schedule is adjusted as follows. 
	Let $\lht$ be the actual load observed at time $t$, and $\alphat$ be the modified storage charge/discharge value at $t$ to ensure feasibility. 

	\textbf{Transformer Constraints}
	\begin{equation}
	\alphat = 0, \quad \text{ if } \ci - \xit - \lht \leq \eta
	\end{equation}
	\begin{equation}
	\alphat = \ci - \lht, \quad \text { if }  \xit \ge \ci - \lht
	\end{equation}
	\textbf{Storage Constraints}
	\begin{equation}
	\alphat = \bi - \sit, \quad \text{ if }  \xit + \sit \ge \bi
	\end{equation}
	\begin{equation}
	\alphat = -\sit, \quad \text{ if }  \xit + \sit \le 0
	\end{equation}
	As indicated earlier, we would like to avoid excessive storage discharging during low energy demand periods.	This constraints is represented as:
	\begin{equation}
	\alphat = -\lht,  \text{ if }  \lht \le -\xit.
	\end{equation}

	\item \textit{The PreDay Algorithm:} This approach uses the previous day's load and emissions factor as input to the linear program to determine the emission-aware schedule. We use a similar approach and modify the schedule as above to ensure that constraints are not violated. 
\end{enumerate}

\section{Experimental Results\label{sec:exp}}

In this section, we evaluate our approach and compare it to the optimal approach and other heuristic approaches. 

\subsection{Load Forecasting Uncertainty}
\label{sec:forecast}
\begin{figure}[t]
\centering
\includegraphics[width=3.in]{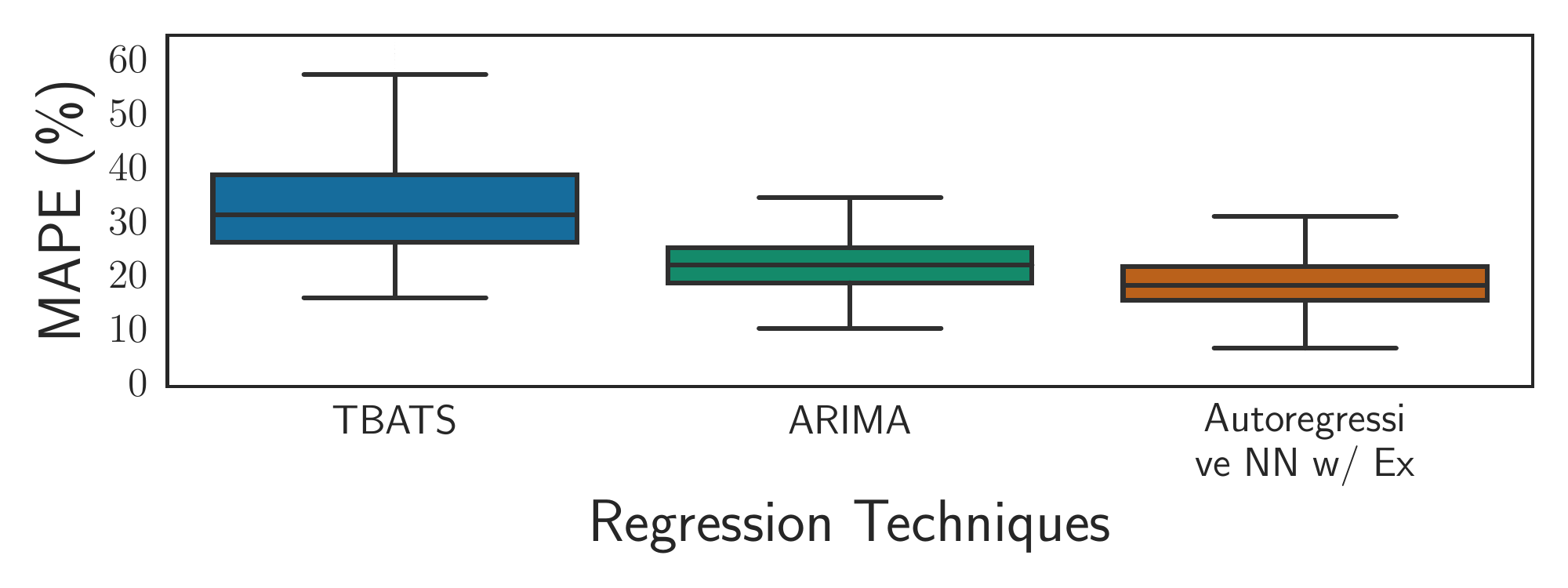}  
\caption{Efficacy of load forecasting methods. The figure shows there can be significant uncertainty in forecasts. }
\label{fig:regression}
\end{figure}

First, we evaluate the efficacy and accuracy of our proposed load forecasting method in Section~\ref{sec:load_forecasting}.
We compare our forecasting method based on state-of-the-art neural network approach with two popular statistical time series techniques --- ARIMA~\cite{brockwell2002introduction} and TBATS~\cite{de2011forecasting}. 
Figure~\ref{fig:regression} compares the performance of the proposed regression technique with the two baseline approaches. 
The results show the distribution of mean absolute percentage error (MAPE) values for load forecast at all transformers evaluated over a period of one year.
Based on our analysis, TBATS has the highest average MAPE of 34.17\%, while the MAPE of ARIMA was 21.5\%. The performance of the autoregressive neural network with exogenous variables outperformed all other techniques and has the lowest average MAPE value of 20.14\%. 
In our experiments, we observed that including exogenous variables improves the accuracy of our forecast significantly.
Despite its higher accuracy, we observe that the forecast
still contains error --- indicating uncertainty in prediction. {\em The presence
of such error is a motivation for leveraging robust optimization for emission-aware
storage scheduling. }

\subsection{Emission-aware Storage Schedule}

\begin{figure}[t]
\centering
\includegraphics[width=3.in]{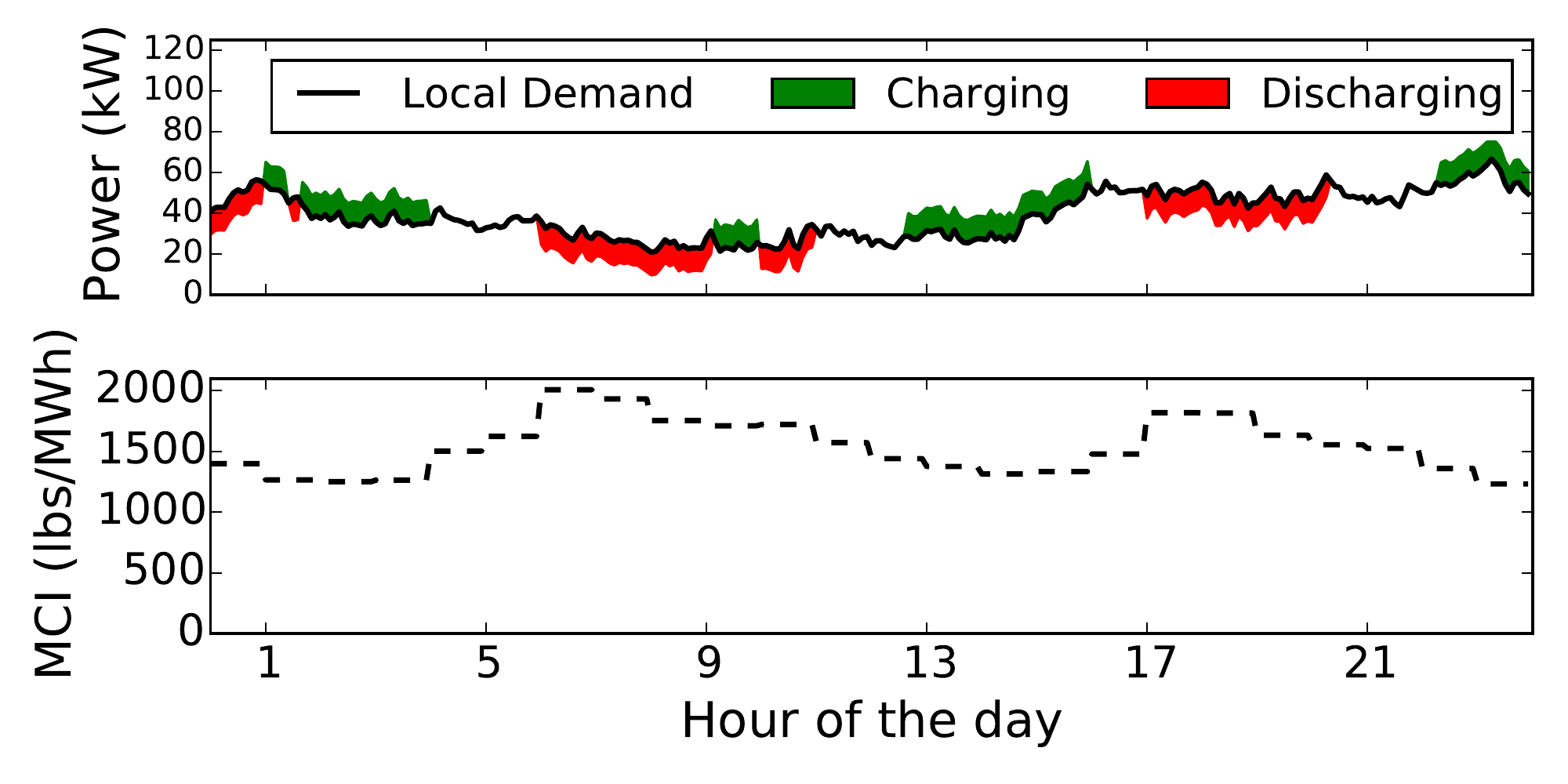}  
\caption{Battery charge and discharge based on our emission-aware energy schedule. Our emission-aware algorithm discharges battery when marginal carbon intensity (MCI) is high.}
\label{fig:illustrative}
\end{figure}

Figure~\ref{fig:illustrative} depicts our emission-aware storage scheduling approach in action. The figure shows the impact of the storage schedule on the load observed at a transformer for a sample day overlayed with the local demand. As shown, discharging action occurs when the marginal emissions are high, e.g., between 6 am to 9 am, which represents the high polluting hours of the day.  Conversely, charging occurs when the marginal emissions are low, usually between 1  to 4 pm. Based on the overall energy usage and the mix of fuels used at different times of the day, the alternating charging and discharging actions at this transformer mitigates 17.5 kg of carbon emissions.\\
{ \em Result: Emissions-aware distributed energy storage has significant potential to reduce carbon emissions at the grid-scale. }

\subsection{Benefits of Emission-aware Scheduling}

\begin{figure}[t]
\centering
\includegraphics[width=2.5in]{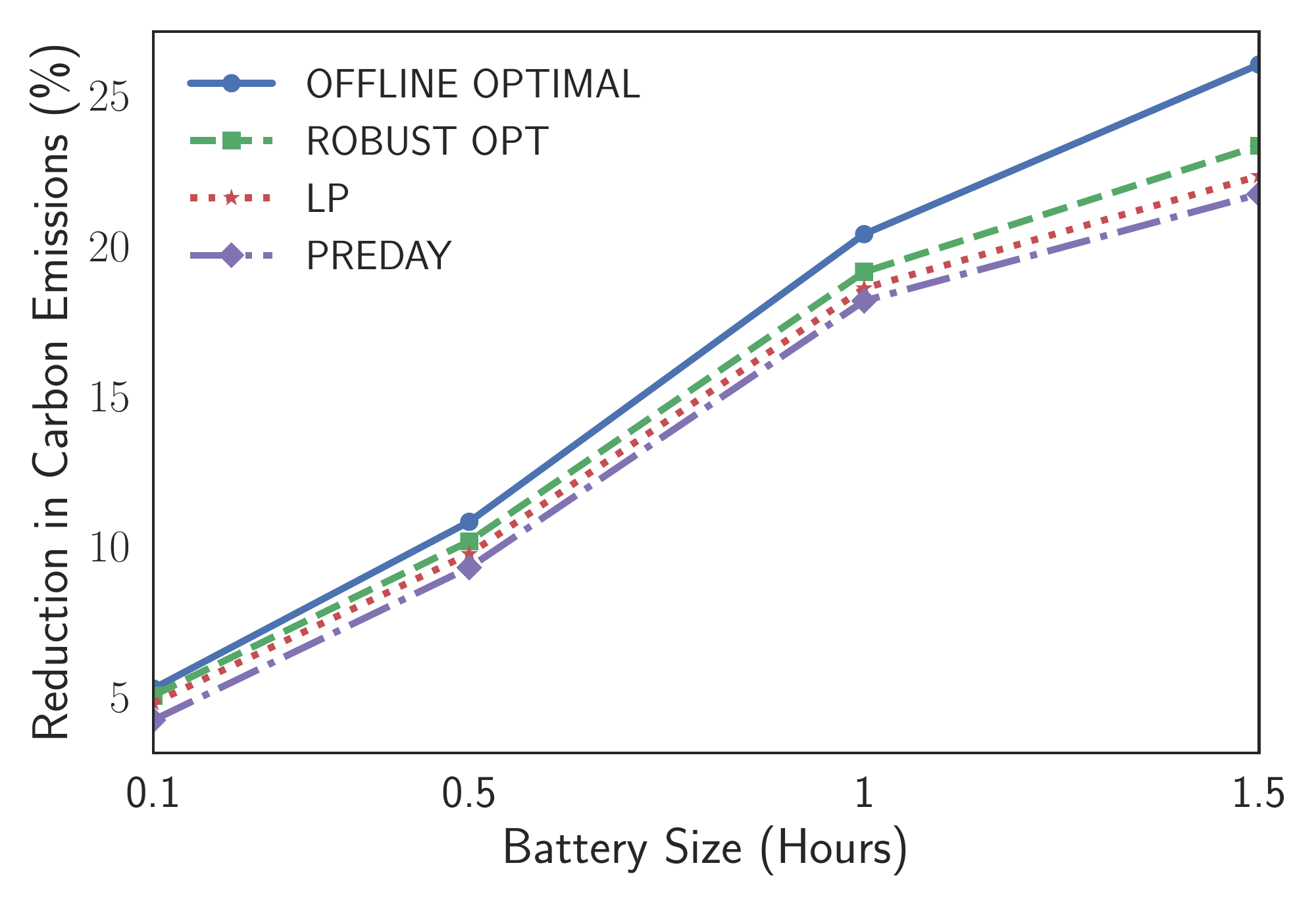}  
\vspace{-0.2cm}
\caption{Carbon emissions reduction for different battery sizes. The battery size is computed as the number of hours it can sustain the maximum load of the transformer. }
\label{fig:batterysize}
\end{figure}

We analyze the change in carbon savings\footnote{We use marginal carbon intensity and change in battery state compared to the previous day to calculate the daily carbon savings.} by varying the size of the energy storage. We size the battery as the number of hours it can support the annual maximum load of the transformer. Thus, one hour of battery capacity indicates that it can support the maximum load of the transformer for an hour. Figure~\ref{fig:batterysize}  shows the reduction in carbon emissions with increasing battery size for different algorithms. 
We observe that the carbon emissions reduces with increasing battery size. This is because a larger battery has more flexibility in shifting transformer load, where batteries can charge during lower emissions and discharge during high carbon emissions period. Even with a battery size of 0.5 hours, we observe that our emissions-aware algorithm achieves 10.16\% reduction in carbon emissions. 
Further, a battery size of 1.5 hours can annually save $>$0.5 million kg of carbon emissions --- equivalent  to 23.3\% reduction in electric grid emissions. 

We also compare our robust optimization with baseline approaches described in Section~\ref{sec:baseline}. 
The optimal approach provides the maximum carbon savings that can be achieved. However, the optimal needs the full information in advance that is not practical. We observe the gap between the optimal and our robust optimization approach is less than $1.2$\% having battery size less than or equal to one hour.\\
{\em Result:  Robust optimization \emph{consistently} performs better than the other baseline approaches; with a 1.5hr battery, it can save $>$0.5 million kg of carbon emissions annually, a 23.3\%  reduction in emissions. }



\subsection{Impact of Storage Parameters}

\begin{figure}[t]
\centering
\includegraphics[width=2.5in]{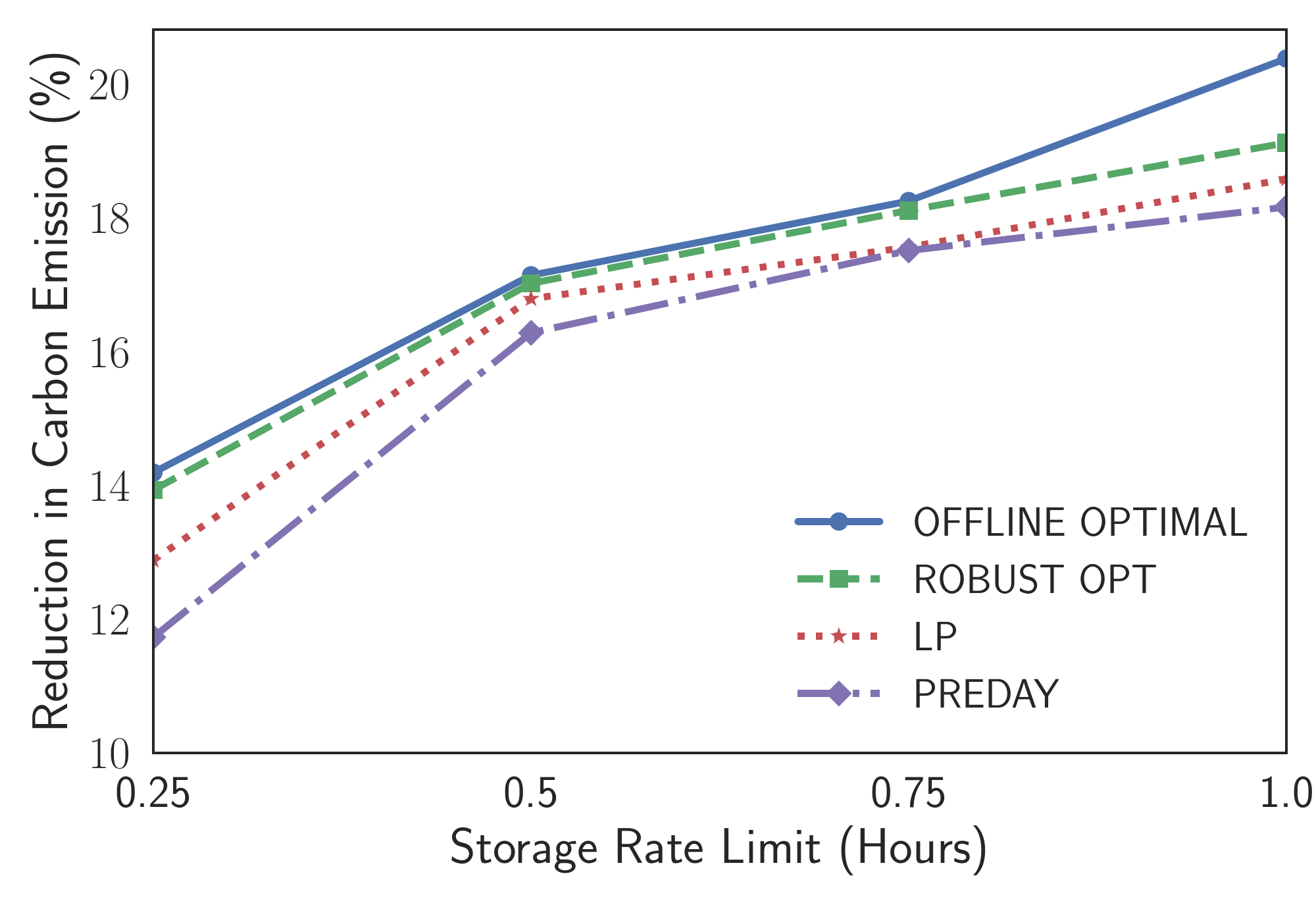}  
\vspace{-0.2cm}
\caption{Carbon emissions reduction for different charge/discharge rate and a battery size of one hour.}
\label{fig:batteryrate}
\end{figure}

Next, we study the effects of different storage parameters---charge and discharge rates---on carbon emission reduction.
  We fix the energy storage size such that it can sustain the maximum load at the transformer for one hour and vary the charge/discharge rate.  
The charge/discharge rate is set such that the fraction indicates the percentage of the maximum load at the transformer the energy storage can charge or discharge. Thus, a 0.25 hour charge/discharge 
rate can discharge at one-fourth the maximum load at the transformer. 
As seen in the Figure~\ref{fig:batteryrate}, with a charge/discharge rate of 0.25 hour, our robust optimization approach achieves a carbon emission reduction of 13.9\%. 
However, an increase in charge/discharge rate further reduces carbon emissions. This is because a higher discharge rate is able to reduce demands thereby minimizing the 
need to utilize generation sources with high emission footprints. In particular, we observe that the reduction in carbon emissions increases by 37.2\% (from 13.93\% to 19.12\%) when the rate is increased from 0.25 to 1 hour.\\
{\em Result: Our RO approach yields 13.9\% emission reductions even at modest battery discharge rates.}

\subsection{Impact of Storage Penetration}

\begin{figure}[t]
\centering
\includegraphics[width=2.5in]{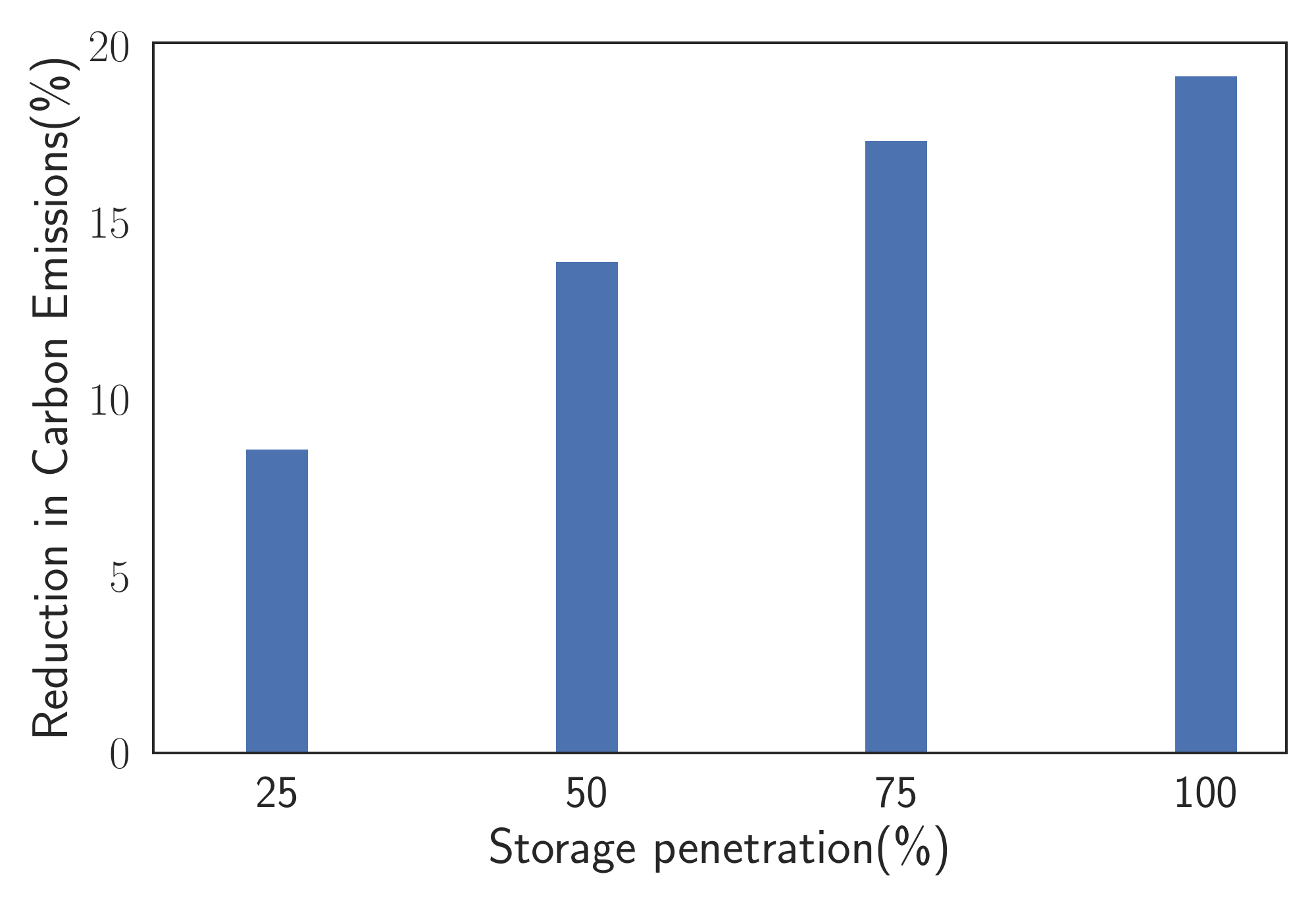}  
\vspace{-0.3cm}
\caption{Carbon emissions reduction for different levels of storage penetration across transformers. A 50\% penetration indicate half the transformers have storage units installed.}
\label{fig:penetration}
\end{figure}

We study the benefit that comes from installing energy storage at only a fraction of the transformers in the grid.  
Like before, in this experiment, we fix the energy storage size such that it can sustain the transformer load at its maximum for one hour 
and select  transformers at random, where batteries are installed. Figure~\ref{fig:penetration} shows the reduction in carbon emissions for different storage penetration levels.
An energy storage penetration of 25\% can achieve 8.5\% carbon emission reduction. However, if 50\% of the transformers install energy storage, the carbon emission reductions 
improves to 13.9\%, a 63.5\% improvement in emissions reduction. This is because higher energy storage penetration can offset more loads that have high emissions footprint. 
Further, if all the transformers have energy storage, the reduction in carbon emission is 19.12\%.\\
{\em Result: Even a modest 1-hr battery can yield up to 19\% reduction in carbon emissions. Larger batteries and higher penetration levels will provide much higher reductions.}

\section{Related Work\label{sec:rel}}

\emph{Energy Storage Systems in the Electric Grid.}
There has been significant work on using energy storage in the electric grid~\cite{irwin2017enabling,hill2012battery}. 
However, the majority of work has focused on improving grid stability or cost arbitrage. Our work focuses on using energy storage to reduce grid carbon emissions. 


Additionally, shifting the energy demand has been suggested in the literature by introducing flexibility in loads through a mechanism called \textit{demand response}~\cite{gnauk2012leveraging,scott2013residential}. 
Monetary incentives are set aside to compensate for the customers participating in demand response.
However, demand response involves customer buy-in and often include installing specialized hardware on the electric loads, which may not always be feasible. On the contrary, grid operators around the world can readily employ our approach by utilizing carbon intensity values from the set of power plants they control.

\emph{Load Forecasting.}
In smart grids research, load forecasting is a widely studied problem. The regression techniques used to solve this problem range from traditional time series approaches such as ARIMA~\cite{nguyen2017short} to neural network~\cite{kong2017short}. Traditionally, grid-level load forecasting was used to assess power systems security, schedule maintenance services, etc. Our regression model produces forecasts at the transformer-level and improves over the state-of-the-art technique~\cite{kong2017short}. 

\emph{Robust Optimization for Scheduling in Smart Grid.}
Robust optimization has been extensively used to solve different problems in different application domains that deal with uncertainty, including smart grid. Some examples are generator placement~\cite{wang2014robust,wang2015stochastic}, EV charging scheduling~\cite{korolko2017robust}, storage sizing~\cite{kazhamiaka2018robust}.
As compared to the other stochastic approaches it has several advantages: (1) it does not require stochastic modeling of uncertain parameters in terms of probability distribution functions; (2) by defining the notion of budget of uncertainty~\cite{bertsimas2004price}, its provides a design space to trade-off between the robustness and performance of the decision making. Note that the notion of uncertainty set has been used in other theoretical approaches such as competitive analysis~\cite{hajiesmaili2016online}, however, the algorithm approach used in~\cite{hajiesmaili2016online} is problem-specific and cannot be applied to our emission-aware scheduling scenario.


\section{Conclusion\label{sec:conc}}

The benefits of distributed energy storage have been previously studied for grid optimizations such as peak shaving, price arbitrage, and demand-response.
However, in this work, we focus on using distributed energy storage to reduce the emission footprint of electricity generation. 
Our main insight is that energy storage can help utility companies reduce the reliance on less efficient and most carbon-intensive power plants, 
shifting electric demand from high polluting periods to low polluting periods. 
We formulated the problem of emission-aware scheduling as an optimization problem with the objective of minimizing the carbon emission, subject to transformer and storage operational constraints. 
Given the dynamics in transformer-level load, we leveraged robust optimization to handle the uncertainty in load predictions. 
We evaluated our emission-aware energy storage scheduling approach on a  dataset containing 100 transformers connected to over 1,340 electric meters in a city in the Northeastern part of the US. 
Our analysis showed that our approach can offset  $>$0.5  million kg in annual carbon emissions, which is equivalent to a 23.3\% reduction in the electric grid emissions. 

\noindent{\bf Acknowledgements} We thank the anonymous reviewers for their helpful comments. This research was supported in part by NSF grants 1645952, 1908298 and MA Department of Energy Resources.

\bibliographystyle{acm}
\raggedright
\bibliography{paper}

\begin{thebibliography}{10}

\bibitem{epri2016}
Wholesale electricity market design initiatives in the united states: Survey
  and research needs.
\newblock {\em EPRI,Technical Results, available at
  \url{https://www.epri.com/pages/product/3002009273/}\/} (2016).

\bibitem{green-mountain}
Green mountain power.
\newblock \url{https://greenmountainpower.com/product/powerwall/}, 2019.

\bibitem{ben2009robust}
{\sc Ben-Tal, A., El~Ghaoui, L., and Nemirovski, A.}
\newblock {\em Robust optimization}, vol.~28.
\newblock Princeton University Press, 2009.

\bibitem{bertsimas2004price}
{\sc Bertsimas, D., and Sim, M.}
\newblock The price of robustness.
\newblock {\em Operations research 52}, 1 (2004), 35--53.

\bibitem{bnef}
{G}lobal {S}torage {M}arket to {D}ouble {S}ix {T}imes by 2030.
\newblock
  \url{https://about.bnef.com/blog/global-storage-market-double-six-times-2030/},
  2017.
\newblock Accessed: 2018-08-08.

\bibitem{Borodin98}
{\sc Borodin, A., and El-Yaniv, R.}
\newblock {\em Online computation and competitive analysis}.
\newblock Cambridge University Press, 1998.

\bibitem{brockwell2002introduction}
{\sc Brockwell, P.~J., Davis, R.~A., and Calder, M.~V.}
\newblock {\em Introduction to time series and forecasting}, vol.~2.
\newblock Springer, 2002.

\bibitem{de2011forecasting}
{\sc De~Livera, A.~M., Hyndman, R.~J., and Snyder, R.~D.}
\newblock Forecasting time series with complex seasonal patterns using
  exponential smoothing.
\newblock {\em Journal of the American Statistical Association 106}, 496
  (2011), 1513--1527.

\bibitem{diaconescu2008use}
{\sc Diaconescu, E.}
\newblock The use of narx neural networks to predict chaotic time series.
\newblock {\em Wseas Transactions on computer research 3}, 3 (2008), 182--191.

\bibitem{eia}
{EIA} {D}ata {B}rowser.
\newblock \url{https://www.eia.gov/electricity/data/browser/}, 2018.
\newblock Accessed: 2018-08-08.

\bibitem{gnauk2012leveraging}
{\sc Gnauk, B., Dannecker, L., and Hahmann, M.}
\newblock Leveraging gamification in demand dispatch systems.
\newblock In {\em Proceedings of the 2012 Joint EDBT/ICDT workshops\/} (2012),
  ACM, pp.~103--110.

\bibitem{hajiesmaili2016online}
{\sc Hajiesmaili, M.~H., Chau, C.-K., Chen, M., and Huang, L.}
\newblock Online microgrid energy generation scheduling revisited: The benefits
  of randomization and interval prediction.
\newblock In {\em Proceedings of the Seventh International Conference on Future
  Energy Systems\/} (2016), ACM.

\bibitem{hill2012battery}
{\sc Hill, C.~A., Such, M.~C., Chen, D., Gonzalez, J., and Grady, W.~M.}
\newblock Battery energy storage for enabling integration of distributed solar
  power generation.
\newblock {\em IEEE Transactions on smart grid 3}, 2 (2012), 850--857.

\bibitem{emissionfactor}
{\sc Hoedl, S.}
\newblock {L}ocation-{B}ased {ISONE} {P}ollutant {I}ntensity {A}nalysis.
\newblock Tech. rep., Harvard University's Office for Sustainability, 08 2016.

\bibitem{irwin2017enabling}
{\sc Irwin, D., Iyengar, S., Lee, S., Mishra, A., Shenoy, P., and Xu, Y.}
\newblock Enabling distributed energy storage by incentivizing small load
  shifts.
\newblock {\em ACM Transactions on Cyber-Physical Systems 1}, 2 (2017), 10.

\bibitem{iso-ne}
{ISO} {N}ew {E}ngland {W}eb {S}ervices {API} v1.1.
\newblock \url{https://webservices.iso-ne.com/docs/v1.1/}, 2018.
\newblock Accessed: 2018-08-08.

\bibitem{iyengar2016analyzing}
{\sc Iyengar, S., Lee, S., Irwin, D., and Shenoy, P.}
\newblock Analyzing energy usage on a city-scale using utility smart meters.
\newblock In {\em Proceedings of the 3rd ACM International Conference on
  Systems for Energy-Efficient Built Environments\/} (2016).

\bibitem{ji2019coordinating}
{\sc Ji, C., Hajiesmaili, M.~H., Gayme, D.~F., and Mallada, E.}
\newblock Coordinating distribution system resources for co-optimized
  participation in energy and ancillary service transmission system markets.
\newblock In {\em Proc. of IEEE American Control Conferences\/} (2019).

\bibitem{kazhamiaka2018robust}
{\sc Kazhamiaka, F., Ghiassi-Farrokhfal, Y., Keshav, S., and Rosenberg, C.}
\newblock Robust and practical approaches for solar pv and storage sizing.
\newblock In {\em Proceedings of the Ninth International Conference on Future
  Energy Systems\/} (2018), ACM, pp.~146--156.

\bibitem{kong2017short}
{\sc Kong, W., Dong, Z.~Y., Jia, Y., Hill, D.~J., Xu, Y., and Zhang, Y.}
\newblock Short-term residential load forecasting based on lstm recurrent
  neural network.
\newblock {\em IEEE Transactions on Smart Grid\/} (2017).

\bibitem{korolko2017robust}
{\sc Korolko, N., and Sahinoglu, Z.}
\newblock Robust optimization of ev charging schedules in unregulated
  electricity markets.
\newblock {\em IEEE Transactions on Smart Grid 8}, 1 (2017), 149--157.

\bibitem{kyriakides2007short}
{\sc Kyriakides, E., and Polycarpou, M.}
\newblock Short term electric load forecasting: A tutorial.
\newblock In {\em Trends in Neural Computation}. Springer, 2007, pp.~391--418.

\bibitem{lee2018vsolar}
{\sc Lee, S., Shenoy, P., Ramamritham, K., and Irwin, D.}
\newblock vsolar: Virtualizing community solar and storage for energy sharing.
\newblock In {\em Proceedings of the Ninth International Conference on Future
  Energy Systems\/} (2018), pp.~178--182.

\bibitem{nguyen2017short}
{\sc Nguyen, H., and Hansen, C.~K.}
\newblock Short-term electricity load forecasting with time series analysis.
\newblock In {\em Prognostics and Health Management (ICPHM), 2017 IEEE
  International Conference on\/} (2017), IEEE, pp.~214--221.

\bibitem{nunna2013energy}
{\sc Nunna, H.~K., and Doolla, S.}
\newblock Energy management in microgrids using demand response and distributed
  storage - a multiagent approach.
\newblock {\em IEEE Transactions on Power Delivery 28}, 2 (2013), 939--947.

\bibitem{RPI-MS-Thesis}
{\sc Olivieri, Z.}
\newblock {Optimization of Residential Battery Energy Storage System Scheduling
  for Cost and Emissions Reductions}.
\newblock {\em RIT Scholar Works\/} (2017).

\bibitem{Padhy04}
{\sc Padhy, N.}
\newblock Unit commitment a bibliographical survey.
\newblock {\em IEEE Trans. On Power Systems 19}, 2 (2004), 1196--1205.

\bibitem{rogers2013evaluation}
{\sc Rogers, M.~M., Wang, Y., Wang, C., McElmurry, S.~P., and Miller, C.~J.}
\newblock Evaluation of a rapid lmp-based approach for calculating marginal
  unit emissions.
\newblock {\em Applied energy 111\/} (2013), 812--820.

\bibitem{scott2013residential}
{\sc Scott, P., Thi{\'e}baux, S., Van Den~Briel, M., and Van~Hentenryck, P.}
\newblock Residential demand response under uncertainty.
\newblock In {\em International Conference on Principles and Practice of
  Constraint Programming\/} (2013), Springer, pp.~645--660.

\bibitem{walawalkar2007economics}
{\sc Walawalkar, R., Apt, J., and Mancini, R.}
\newblock Economics of electric energy storage for energy arbitrage and
  regulation in new york.
\newblock {\em Energy Policy 35}, 4 (2007), 2558--2568.

\bibitem{wang2015stochastic}
{\sc Wang, Z., Chen, B., Wang, J., and Begovic, M.~M.}
\newblock Stochastic {DG} placement for conservation voltage reduction based on
  multiple replications procedure.
\newblock {\em IEEE Transactions on Power Delivery 30}, 3 (2015), 1039--1047.

\bibitem{wang2014robust}
{\sc Wang, Z., Chen, B., Wang, J., Kim, J., and Begovic, M.~M.}
\newblock Robust optimization based optimal {DG} placement in microgrids.
\newblock {\em IEEE Transactions on Smart Grid 5}, 5 (2014), 2173--2182.

\bibitem{zhang2018peak}
{\sc Zhang, Y., Hajiesmaili, M.~H., Cai, S., Chen, M., and Zhu, Q.}
\newblock Peak-aware online economic dispatching for microgrids.
\newblock {\em IEEE Transactions on Smart Grid 9}, 1 (2018), 323--335.

\end{thebibliography}

\end{document}